\documentclass[aps,prl,superscriptaddress,twocolumn,showpacs]{revtex4}

\usepackage[pdftex]{graphicx}
\usepackage{dcolumn}
\usepackage{bm}

\begin{document}
\preprint{APS/123-QED}
\title{A new class of organic molecular magnets}

\author{William A. Shelton} 
\affiliation{Oak Ridge National Laboratory, Oak Ridge, Tennessee 37831, USA.}
\email{sheltonwajr@ornl.gov}
\homepage{http://www.csm.ornl.gov/ccsg/html/staff/shelton.html}
\author{Edoardo Apr\`{a}}
\affiliation{Oak Ridge National Laboratory, Oak Ridge, Tennessee 37831, USA.}
\author{Bobby G. Sumpter}
\affiliation{Oak Ridge National Laboratory, Oak Ridge, Tennessee 37831, USA.}
\author{Aldilene Saraiva-Souza} 
\affiliation{Departamento de F\'isica, Universidade Federal do Cear\'a, 60455-900, Fortaleza, 
Cear\'a, Brazil.}
\author{Antonio G. Souza Filho} 
\affiliation{Departamento de F\'isica, Universidade Federal do Cear\'a, 60455-900, Fortaleza, 
Cear\'a, Brazil.}
\author{Jordan Del Nero}
\affiliation{Departamento de F\'isica, Universidade Federal do Par\'a, 66075-110, Bel\'em, 
Par\'a, Brazil.}
\author{Vincent Meunier}
\affiliation{Oak Ridge National Laboratory, Oak Ridge, Tennessee 37831, USA.}

\date{\today}

\begin{abstract}
Using detailed first principles calculations, we have found a new class of stable organic molecular magnets based on zwitterionic molecules possessing donor, $\pi$ bridge, and acceptor groups. These $\it molecules$ are organic molecules containing only C, H and N.  The quantum mechanical nature of the magnetic properties originates from the conjugated $ \pi $ bridge (involving only p electrons) where the exchange interactions between electron spin are relatively strong and local and are independent of the length of the $\pi$ bridge, enabling the easy construction of a molecular magnetic device with specified length. 
\end{abstract}

\maketitle

\begin{figure}
\includegraphics[scale=0.75]{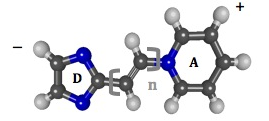}
\caption{\label{fig:betart} A zwitterionic betaine molcule with an imidazol donor (D) and a 
pyridine acceptor (A) separated by alternating C$=$C bridge.}
\end{figure}

\begin{figure}
\includegraphics[scale=0.35]{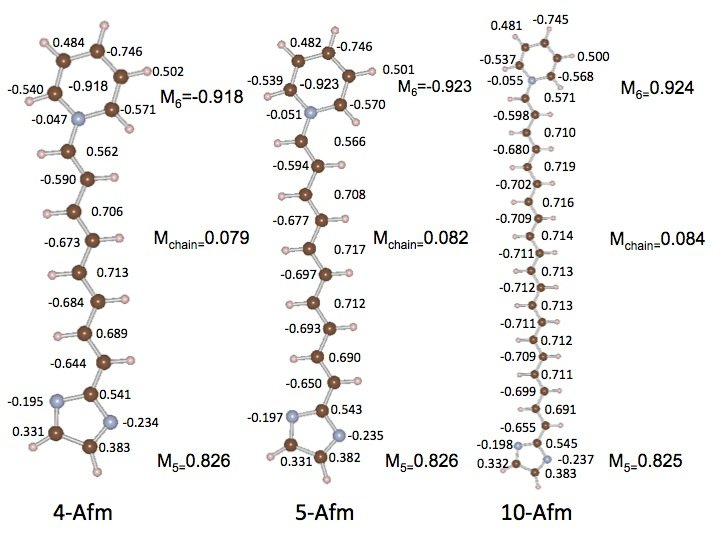}
\caption{\label{fig:CHmag} The magnetic moments for C and N along with the 5,6 membered ring and
C$_2$H$_2$ $\pi$ bridge magnetic moment contributions for chain lengths of 4,5 and 10.} 
\end{figure}

\begin{figure}
\includegraphics[scale=0.50]{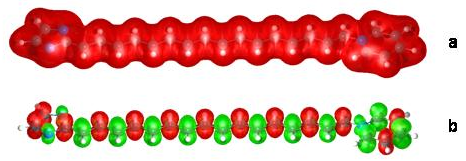}
\caption{\label{fig:spind} Figure 3a shows the delocalization of the charge density.  Figure 3b 
displays the spin density showing the strong Heisenberg character.} 
\end{figure}

\begin{figure}
\includegraphics[scale=0.33]{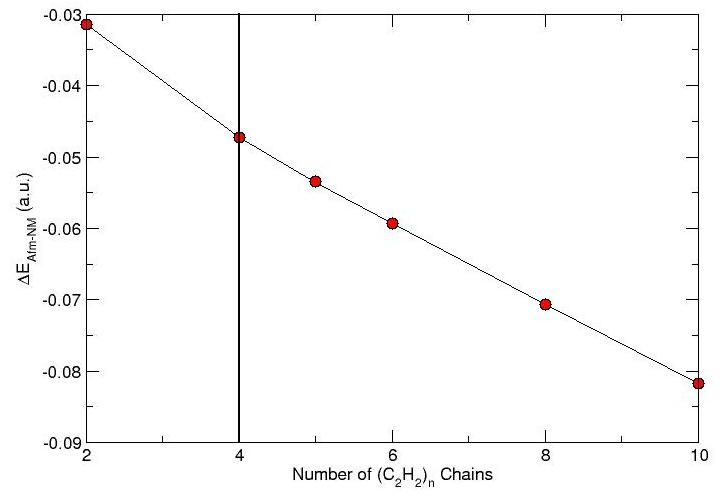}
\caption{\label{fig:EvsCH} $\Delta$E$_{Afm-NM}$ vs C$_2$H$_2$ $\pi$ bridge length.} 
\end{figure}

Metal-free molecular magnets offer intriguing benefits in terms of cost, weight, and diversity in
magnetic properties and ease in processing and synthesis. 
However most traditional molecular-based
magnets are crystals composed of molecular units that contain a variety of metals or
radicals\cite{Blundell2001} that have either a low Curie temperature (T$_C$) or are unstable due 
to the processing techniques used to generate them. 
In the present paper we show that a particular class of organic $\it molecules$, containing
no radicals or metals, can have a stable magnetic ground-state and thus, should have a 
high N\'{e}el temperature (T$_N$).  Moreover, by simply substituting a N with a C atom on 
the 6-membered ring comprising the betaine molecule (see figure~\ref{fig:betart})
the magnetic state can be changed from
antiferrimagntic (Afm) to ferrimagnetic (fm) and it should have a high 
T$_C$.  Magnetic materials that are purely organic molecules 
are uncommon since the p-electrons in carbon based materials generally participate in covalent 
bonds\cite{Makarova2006}.
Our findings indicate that betaine derivatives promise to offer a new class of
molecular magnetic materials with diverse potential applications including organic spintronics, 
information storage and nanoscale sensors.
\linebreak\indent A limited number of molecular-based magnets have been reported\cite{Blundell2001}
and these are typically
composed of organic radicals\cite{M.Tamura1991}  or mixed coordination compounds containing bridging
organic radicals, Prussian Blue type compounds, or charge transfer complexes\cite{Miller2000}. 
For the relatively few organic materials that possess 
magnetic properties, the Curie temperature (T$_C$) is generally very low ($<$36 K)
\cite{M.Tamura1991}\cite{Kroto1985}\cite{Palacio1997} while those with high T$_C$ typically contain 
some metallic atoms\cite{Makarova2001}\cite{Andriotis2003}\cite{Chan2004}\cite{Makarova2006} with 
the organic molecular part of the crystal acting as exchange pathways.  Other allotropes of carbon 
such as carbon nanofoams\cite{Rode2004} and boron-nitride nanotubes\cite{Wu2005} have also been 
reported to exhibit magnetic properties, however either the magnetic properties are very short-lived 
or they depend on doping and defects to achieve a magnetic state where the stability and strength of 
the magnetism is low since the achieveable doping/defect concentration is quite low\cite{Miller1986}.
Organic free radicals are natural candidates for magnetic
materials, but very few are stable enough to be used as magnetic devices.  
\linebreak\indent While it is clear that some progress towards an organic magnetic material has 
been made (in particular for graphitic systems), room temperature, air stable and synthetically 
viable molecules
are still lacking.  Here we report on a zwitterionic molecular system that is composed entirely of C, 
N and H and has a magnetic ground-state.   The betaine derivatives are zwitterions with a 
$\pi$-conjugated structure whose variable $\pi$ bridge length separates the donor and acceptor 
groups\cite{Saraiva-Souza2008}.   $\pi$-conjugated organic molecules are in general good candidates 
for exhibiting magnetic order,\cite{Ovchinnikov1978} although as discussed above, only a relatively 
few successful examples have been reported and those have very low T$_C$.  The betaine
derivatives are unique because they also have an intrinsic  donor and $\pi$ acceptor at different 
ends of the molecule.  As will be shown in this paper, the magnetic properties of these systems are
independent of the bridge length. 
\linebreak\indent The molecular structure for the betaine derivatives is shown in
Figure~\ref{fig:betart} . 
These systems are zwitterions, containing a pyridine acceptor (cation) and imidazole donor (anion)
separated by a variable length $\pi$-conjugated (C$_2$H$_2$)$_n$  bridge.  In this paper, we utilize 
a combination of $\it ab\, initio$ techniques for performing fully relaxed total energy calculations 
using Hartree-Fock (HF), 
complete active space self consistent field (CASSCF), and density functional theory (DFT) within the
framework of the NWChem package\cite{Kendall2000}, to 
determine the intrinsic character of the ground state for several betaine derivatives. 
The standard Pople basis set, 6-311G$^{**}$ was used for all
calculations, yielding similar results. A hybrid approach was chosen since it is known 
that using standard density functional theory methods for purely organic $\pi$ conjugated systems often
fail due to the delocalization of the system\cite{Hirata1998}, and 2) an accurate description of
electron exchange is necessary for determining magnetic properties.  
We therefore only report the DFT results obtained using 100$\%$ HFX with Perdew-Zunger (PZ)
correlation\cite{Perdew1981}.  We note that one can use either PZ or 
Perdew-Burke-Ernzerhof (PBE)\cite{Perdew1996} correlation without affecting the magnetic properties
since the mechanism responsible for antiferrimagnetism 
or ferrimagnetism is purely local.  To determine the effects of non-dynamical correlation on the
magnetic properties, we used CASSCF and the 6-311G$^{**}$ basis set with an active space of up to 
14 active electrons and 14 orbitals.  The CASSCF results are in good agreement with those 
obtained from HFX-DFT indicating that the HFX-DFT approach is adequate in describing the correlation 
in this system. 
\linebreak\indent For the ferromagnetic (FM) state several multiplicities were
considered and full geometry optimization was performed where the triplet 
state was found to be the lowest in energy for the chain lengths considered.  
We note that the initial antiferromagnetic state relaxes to an
antiferrimagnetic state (Afm), and this is the lowest energy state for all bridge lengths examined.
Similarly, the initial FM state relaxes to a ferrimagnetic state (fm) for all bridge lengths considered.
The results 
of our investigation demonstrate that the ground state is magnetic, implying that this class of
molecules are excellent candidates for the development of new organic magnetic systems. 
\linebreak\indent Restricted open-shell Hartree-Fock (ROHF) calculations for the fm type states 
were used to rule out
potential spin-contamination problems. Higher multiplicity calculations for the fm state were also
performed to determine the lowest energy spin state.  For the fm type calculations ROHF, UHF and HFX all
obtained the triplet state as the ground state spin configuration and no significant effects 
associated with spin contamination were found.
\linebreak\indent Table 1 lists the results obtained from geometry optimized calculations using UHF 
and HFX-PZ. Both 
yield the same spin configuration ordering with the lowest energy state being antiferrimagnetic (Afm),
followed by a ferrimagnetic state (fm) with the highest energy state being
non-magnetic (NM).  The Afm-fm energy differences ($\Delta$E$_{Afm-fm}$) listed in table 1 indicate
that the Afm ground-state spin structure is exceptionally stable, exhibiting an equivalent temperature 
that is well above room temperature.  Furthermore, the $\Delta$E$_{Afm-fm}$ is relatively constant as
compared to either the $\Delta$E$_{Afm-NM}$ or $\Delta$E$_{fm-NM}$ energy differences. This is about 
an order of magnitude larger than that typically reported for graphene nanoribbons
\cite{Jiang2007a}\cite{Pisani2007}.  
\begin{table}
\caption{\label{tab:tab1}Total energy difference for fully relaxed UHF and HFX-PZ (labeled PZ)
simulations for NM, fm and Afm states along with the Afm-NM and Afm-NM and fm-NM total enery 
differences for UHF and PZ respectively.  Atomic units are used and $\it n$ in Bet{$\it n$}, defines 
the bridge length.} 
\begin{ruledtabular}
\begin{tabular}{lccccc}
 &UHF &$\Delta$E$_{\rm Afm-fm}$ &PZ&$\Delta$E$_{\rm Afm-fm}$ &$\Delta$E$_{\rm fm-NM}$\\
\hline
$n=2$&&&&&\\
nm & -624.064 &  & -630.792 & &  \\
fm & -624.105 &  & -630.811 & &-0.0187  \\
Afm & -624.117 & -0.0118 & -630.824 &-0.0127  \\
\hline
$n=4$&&&&&\\
nm & -777.843 &  & -786.352 & &   \\
fm & -777.919 &  & -786.387 & & -0.0350  \\
Afm & -777.921 & -0.0114 & -786.399 & -0.0127  & \\
\hline
$n=5$&&&&&\\
nm & -854,734 &  & -864.1333 & & \\
fm & -854.812 &  & -864.1744 & & -0.0411  \\
Afm & -854.823 & -0.0113 & -864.187 & -0.01234  & \\
\hline
$n=6$&&&&&\\
nm & -931.626 &  & -941,916 & & \\
fm & -931.713 &  & -941.963 & & -0.0470  \\
Afm & -931.726 & -0.0130 & -941.975 & -0.0124  \\
\hline
$n=8$&&&&&\\
nm & -1085.410 &  & -1097.481 & & \\
fm & -1085.519 &  & -1097.539 & & -0.0583  \\
Af & -1085.530 & -0.0113 & -1097.551 & -0.0124  \\
\hline
$n=10$&&&&&\\
nm& -1239.195 &  & -1253.046 & & \\
fm& -1239.322 &  & -1253.116 & & -0.0693  \\
Afm& -1239.335 & -0.0130 & -1253.128 & -0.0124
\end{tabular}
\end{ruledtabular}
\end{table}
\linebreak\indent Figure~\ref{fig:CHmag} shows the results for 3 different betaine molecules in 
the Afm and 
fm states with 
bridge lengths of 4, 5, and 10 (see figure~\ref{fig:betart} for a description of a chain length). 
The magnetic 
moments for each C and N site and the total magnetic moments for the chain, 5 and 6 member rings 
are also displayed (the magnetic moments were calculated using a Mulliken population 
analysis\cite{Mulliken1955}).  
For both magnetic states, the magnetic moments of the rings and chains as a function of increasing 
chain lengths are constant.  Interestingly, independent 
of the magnetic state the chains and rings exhibit an Afm and fm type ordering respectively.  For 
the Afm state, the contributions from the rings approximately cancel each other out, while for the 
fm state, only the magnetic moment of the 6-membered ring changes and it varies by $\sim 1 \mu$B.  
These results are the same for all chain lengths investigated in this work.   Thus, it appears that the
magnetism of the system can be viewed as consisting of three nearly independent components, two 
ring structures and an unsaturated chain.
\linebreak\indent The above results imply that the chain length component acts as a strongly 
correlated Afm insulator. 
This can be seen in the spin density plot shown in figure~\ref{fig:spind}b.  Note, that the spin 
density along
the chain contains significant localized Heisenberg spin type character and does not contain any 
spin dimer like qualities,\cite{Rokhasr1988}\cite{Fazekas1974}\cite{Anderson1973} while the charge
density in figure~\ref{fig:spind}a exhibits strong $\pi$-conjugation
throughout the molecule. 
The only fm like coupling occurs in the 5 member ring, which by symmetry 
must occur.  To investigate the magnetic interaction further, a plot of the Afm-NM energy difference 
($\Delta$E$_{Afm-NM}$) vs C$_2$H$_2$ chain length is shown in figure~\ref{fig:EvsCH}.  
For chain lengths of 4 to 10, 
$\Delta$E$_{Afm-NM}$ decreases linearly with increasing chain length, indicating a strong Heisenberg
like behavior with the change in energy being
$\propto -J\sum_{\langle ij \rangle}S_i\cdot S_j$, where the summation is restricted to 
nearest-neighbors.
From the spin density and the magnetic moments in
figure~\ref{fig:CHmag} it is clear that the two rings only
interact locally with the first couple of carbon atoms in the chains, which is why there is a slight
change in linearity for the chain length of 2.  The linear behavior of $\Delta$E$_{Afm-NM}$ with
increasing chain length also indicates it is stable and that the chain acts to insulate the two
ring structures.  This type of screening would account for the change of 1 $\mu$B on the 6-membered 
ring in the fm state.  This can be seen from figure~\ref{fig:CHmag}, where the moments of the two 
rings are
essentially constant with increasing chain length.   We also decomposed the eigenvalues (orbital 
and spin) onto atomic sites for the various chain lengths.  The energy for the highest occupied 
states for the different chain lengths is $\sim$ -0.07 a.u. and essentially constant with increasing 
chain length.  Furthermore, these states are of p$_z$ character and a strong p$_z$ interaction 
occurs between neighboring spin-up and spin-down states within the chain.  This feature taken along 
with figure 2 accurately illustrates the strong localized Heisenberg spin character.   It is 
also important to note that geometric defects in the chains, such as $\it cis$ conformations, do not 
disrupt the Afm ordering (results are not shown). While substitutional defects that 
disrupt the $\pi$ conjugation on the bridge, such as saturation, will destroy the magnetic character.
\linebreak\indent The Afm state in this system is different from Andersons theory of 
superexchange\cite{Anderson1959}   where two 
next-nearest neighbor transition metal atoms interact through a diamagnetic anion.
In addition, C and N atoms have their respective p$_x$ and p$_y$ orbitals
shifted down in energy with respect to the p$_z$ orbitals.  
On the other hand, 
the p$_z$ spin orbitals are symmetry split, such that, the spin alternates between carbon sites on the
chain.  This ordering combined with the local nature of the spins lowers the Coulomb energy between
sites.  The $\pi$-conjugation in the rings is strictly local to the individual rings and is maintained
through primarily nearest-neighbor interactions with the chain. Interactions with neighbors further 
down the chain decrease almost exponentially, producing a short-range effective screening, which 
again highlights the role of the chains to insulate any interactions between the rings, even for 
short chain lengths.  
\linebreak\indent Additional support for the existence of a magnetic ground state was obtained 
from CASSCF calculations using various sized active spaces, ranging from 10 electrons and 10 orbitals 
to 14 electrons and 14 orbitals.  From these calculations it is clear that the natural orbitals show
significant partial occupancies, indicating a noteworthy multi-configurational character of the 
singlet ground state.  
For Bet6 with an active space of (12,12) the unpaired occupancies are: 
1.9448, 1.9332, 1.9242, 1.8799, 1.8570, 1.7075, 0.2933, 0.1525, 0.1202, 0.0739, 0.0523, 0.0613.  
The unpaired electron density is localized on each C site across the C=C bridge and on the donor 
and acceptor rings of the molecule. 
For the larger betaines 7-10, a similar result is found.  
Larger active space calculations did not change the fact that there is an open shell singlet 
ground state.    The triplet state is higher in energy 
than the Afm singlet but below the non-magnetic singlet state.  These results are consistent with 
those given in table 1 for the HFX-DFT and UHF calculations.
\linebreak\indent In this paper we have presented a detailed examination of the fundamental 
electronic structure of a 
new class of zwitterionic molecules based on betaine derivatives.  The results show that these 
compounds have a surprisingly strong local Heisenberg type magnetic character that originates from 
the conjugated  bridge in cooperation with the donor-acceptor character 
of the molecule.  The localized behavior found in this system is compatible with the 
$\pi$-conjugation of the components making up the system, thus maintaining the resonant behavior in 
the systems charge density aiding in its overall stabilization.  The Afm state further stabilizes 
the system because the Coulomb energy is diminished by alternating the spins throughout most of the
system.   Thus, the stabilization of the system is achieved by the resonating nature of the charge
density (kinetic energy) and the localization of the Afm state that results in a lowering of the 
Coulomb energy.   The Afm type ordering found on the conjugated  bridge is critical in establishing 
the short-range local interactions with the donor and acceptor ring structures and the resulting lack 
of long-range interaction between them.  These features are responsible for the stable magnetic
configuration as a function of bridge length and could possibly enable the development of a variety 
of magnetic devices.  Furthermore the magnetic character is not disrupted by conformational disorder.
Stable room temperature molecular magnetic materials that are purely organic are rather rare and the
betaine derivatives promise to offer a new class of molecular magnetic materials with diverse 
potential applications including organic spintronics,  information storage and nanoscale sensors.  
\linebreak\indent This research was supported in part by the Division of Materials Science and Engineering, U.S. Department of Energy and the Center for Nanophase Materials Sciences (CNMS), sponsored by the Division of Scientific User Facilities, U. S. Department of Energy. AS-S is grateful a CNPq fellowship. AGSF acknowledges the FUNCAP and CNPq agencies. AGSF and JDN acknowledge the Rede Nanotubos de Carbono/CNPq and the FAPESPA agency. 
 
\bibliography{shelton}

\end{document}